# FABRICATION OF MESSAGE DIGEST TO AUTHENTICATE AUDIO SIGNALS WITH ALTERNATION OF COEFFICIENTS OF HARMONICS IN MULTI-STAGES (MDAC)

Uttam Kr. Mondal[1] and J.K.Mandal[2]

[1]Dept. of CSE & IT, College of Engg. & Management, Kolaghat, W.B, India
`uttam_ku_82@yahoo.co.in`
[2]Dept. of CSE, University of Kalyani, Nadia (W.B.), India
`jkm.cse@gmail.com`

## ABSTRACT

*Providing security to audio songs for maintaining its intellectual property right (IPR) is one of chanllenging fields in commercial world especially in creative industry. In this paper, an effective approach has been incorporated to fabricate authentication of audio song through application of message digest method with alternation of coefficients of harmonics in multi-stages of higher frequency domain without affecting its audible quality. Decomposing constituent frequency components of song signal using Fourier transform with generating secret code via applying message digest followed by alternating coefficients of specific harmonics in multi-stages generates a secret code and this unique code is utilized to detect the originality of the song. A comparative study has been made with similar existing techniques and experimental results are also supported with mathematical formula based on Microsoft WAVE (".wav") stereo sound file.*

## 1. INTRODUCTION

Today's creative organizations are facing competitive market for spreading business as well as holding goodwill. Creating a quality product involved a lot of investment as well as money. People are finding easier way to put less effort or investing money and producing product for existence in this contemporary market. Some of them are applying technology for making piracy of original versions and producing lower price products. This intension is a frequent phenomenon for digital audio/video industry with improvement of digital editing technology [4]. Even, it is quite harder to listeners to find the original from pirated versions. Therefore, it is a big challenge for business persons, computer professionals or other concern people to ensure the security criteria of originality of songs[1, 2] and protect from the release of duplicate versions.

In this paper, a framework for identifying a particular song with the help of unique secret code obtained through coefficient alternation of harmonics in multi-levels over the song signal without affecting its quality has been presented. Proposed technique is evolved by decomposing frequency components of the signal and alternating coefficients of specified near harmonics in higher frequency region by generating a secret code. Embedded song signals with secrete code can easily distinguish the original from similar available songs. It is experimentally observed that alternating coefficients of harmonics will not affect the song quality but provide a level of security to protect the piracy of song signal.

Organization of the paper is as follows. Embedding secret key and coefficient alternation are presented in section 2.1 and 2.2 respectively. The authentication procedure has been depicted in section 2.3 that of extraction in section 2.4. The separation of embedding message is performed in section 2.5. Experimental results are given in section 3. Conclusions are drawn in section 4. References are given at end.

## 2. THE TECHNIQUE

The scheme fabricates the secret key with help of message digest method followed by alteration of some of the coefficients in higher frequencies [beyond audible range]. Algorithms namely MDAC-FSK and MDAC-MCAL are proposed as double security measure, the details of which are given in section 2.1 and section 2.2 respectively.

### 2.1. Fabricating Secret Key (MDAC - FSK)

Embedding a secret key in the specific positions of signal is computed through a hash function. The procedure of embedding secret key is depicted in the following algorithm.

*Algorithm:*

*Input:*   Original song signal, text message with selected cryptographic hash function.
*Output:*  Modified song signal with embedded message.
*Method:* Embedding a secret key in the lower frequency areas to avoid distortion of the quality of songs as follows

Step 1:   Find all frequency components which are less than 300 Hz using Fast Fourier Transform (FFT).
Step 2:   Take a secret information/key and apply cryptographic hash function MD5 (Message Digest algorithm) to produce a 128-bit (16-byte) hash value (fixed-size bit string) [using steps 3 to 6, each character will embed within a set of four frequency components (in sequence) which are less than 300Hz (step 1). Therefore, half of the range of frequency components (1-150 Hz) will be used to embed at most 37 characters].
Step 3:   Choose a song identification message (secret key) and find its hash value (step 2) in ASCII bit pattern. Suppose, the secret message is "The quick brown fox jumps over the lazy dog", its equivalent hash value, i.e., MD5 ("The quick brown fox jumps over the lazy dog") = 9e107d9d372bb6826bd81d3542a419d6, bit sequence in binary is 10011110000100000111110110011101001101110010101110110… respectively, i.e., a stream of 16 bytes.
Step 4:   Divide the bit sequence of secret key into small bit patterns of size two i.e. total number of small bit patterns is (16*8)/2=64. To append 64 bit pairs in the song signal, need at least 64 rows in sampled data set.
Step 5:   Represent each bit pair into equivalent lower magnitude value in sequence. (00 will be represented by 0, 01 by 0.0001,10 by 0.0010, and 11 by 0.0011).
Step 6:   Put the lower magnitude values over the sampled data of the signal using the following rules.

   i.   *Choose a position starting from first position up to 300 Hz frequency range [if message size is less than half of frequency range (1-150Hz) then, find suitable*

> *gap between appended positions]. Here, message size 64 (after taking 2 bit units), therefore, require gap is 150/64 or, 1 position, i.e., each value of message should add at $1^{st}$, $2^{nd}$, $3^{rd}$ ,… position and make frequency component zero at all these specified positions of two columns of sampled data set .*

> ii. *Append the magnitude values (in sequence) in specified positions of song signal by following method.*
> $i^{th}$ *position value(ival) of message should add at $k^{th}$ position of x, i.e., x(k,1)=ival and x(k,2)= ival, where k = (i-1)*2 + 1. If $i^{th}$ position is the last position then, x(k,1)=ival .*

> iii. *Apply same method starting from middle position of alternate channel as the initial appended channel as above (step ii) for embedding same information to verify the unchanged magnitude value during retrieval.*

> iv. *Stop when all magnitude values are assigned to their respective positions over the specified frequency range of the signal.*

Step 7:   Apply inverse FFT to get back the sampled values of modified song signal.

Therefore, if any value altered in processing, it will create a difference with the assigned magnitude values which present throughout the signal and changing one position will change the content of embedded message.

## 2.2. Multi-stage Coefficient Alternation (MDAC - MCAL)

Consider figure 1 where three harmonics $k^{th}$, $(k+p)^{th}$ and $(k+q)^{th}$ are alternating coefficients in specified positions,i.e., $n^{th}$ with $(n+x)^{th}$ position,$(n+2x)^{th}$ with $(n+3x)^{th}$ position, etc where, x is the half of the sine wave length that uses as for selecting distances among harmonics for exchanging coefficient values. Alternating coefficient values can be expressed by following mathematical formula.

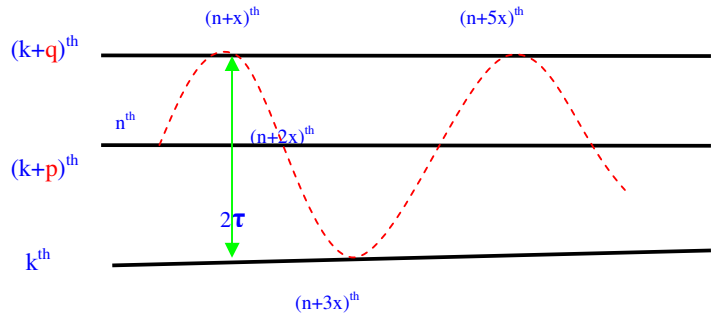

Figure 1.  Selected harmonics

Let, x(n,2) is set of total sampled data of a song, Fourier series of a function $f(x)$ can be written as

$$f(x) = \frac{1}{2}a_0 + \sum_{n=1}^{\infty} a_n \cos(nx) + \sum_{n=1}^{\infty} b_n \sin(nx),$$

Where,

$$a_0 = \frac{1}{\pi} \int_{-\pi}^{\pi} f(x) dx$$

$$a_n = \frac{1}{\pi} \int_{-\pi}^{\pi} f(x) \cos(nx) dx$$

$$b_n = \frac{1}{\pi} \int_{-\pi}^{\pi} f(x) \cos(nx) dx \quad \text{and } n=1, 2, 3.....$$

The Fourier coefficients ($a_n$, $b_n$) are commonly expressed using the formulae which is given in equation (1).

$$c_n = \frac{1}{2\pi} \int_{-\pi}^{\pi} f(x) e^{-inx} dx. \tag{1}$$

The Fourier coefficients $a_n, b_n, c_n$ are related via

$$a_n = c_n + c_{-n} \text{ for } n = 0, 1, 2, \ldots,$$

and
$$b_n = i(c_n - c_{-n}) \text{ for } n = 1, 2, \ldots$$

We can also use Euler's formula,
$$e^{inx} = \cos(nx) + i\sin(nx),$$

Where *i* is the imaginary unit, to give a more concise formula:

$$f(x) = \sum_{n=-\infty}^{\infty} c_n e^{inx}.$$

The notion of a Fourier series can also be extended to a customize form by modifying *f*, such as $F$ or $\hat{f}$, and functional notation often replaces subscripting, which is given in equation (2).

$$f(x) = \sum_{n=-\infty}^{\infty} \hat{f}(n) \cdot e^{inx}$$
$$= \sum_{n=-\infty}^{\infty} F[n] \cdot e^{inx} \tag{2}$$

Particularly when the variable *x* represents time, the coefficient sequence is a frequency domain representation (discrete).

It has been experimentally observed the alternation of the coefficients of closely harmonics of song signal is not changing the audio quality of song. Therefore, finding coefficient values of closely specified harmonics and interchanging those coefficients, we can derived another song signal which will carry some authenticate information with the modified song signal without affecting its audible quality. $C_n$ of equation (1) can be represented by equation (3).

$$C_n = \sum_{k=0}^{N-1} f(x) e^{-i2\pi nk/N} \qquad (3)$$

i.e., $\mathbf{C_n}$ will be interchanged with $\mathbf{C'_{n+px}}$ coefficient. The value of p and x (half of sine wave length) is determined based on quality of song.

### 2.3. Authentication

The authentication algorithm of the proposed technique is given below.

Algorithm:

*Input:*   Modified song signal with embedded message.
*Output:*  Modified song signal with interchanged coefficients of specified harmonics in multi-stages.
*Method:* Extraction of coefficient values and altering of the same using a hash function without affecting quality of the song.

Step 1:  Apply FFT over input signal (song) x(n,2) to find the magnitude values of frequencies of the song signal and put into another array s(i) , 1<=i<=n. Obtain the coefficients($C_i$) using equation (3) of some specified harmonics.

Step 2:  Interchange the values of coefficients by designing a path moving as a waving as sine wave, where wavelength is 2x, wave height is 2$\tau$ and origin point of sine wave is the point above 20,000 Hz [beyond audible range] with closely harmonics (for selected p and q [figure 1]) between two values of ix, where i=1, 2, 3, … and i<=n [using equation (3) and step 2].

Step 3:  Apply IFFT over the modified values of s(i), 1<=i<=n to get the modified authenticated song signal.

### 2.4. Extraction

The decoding is performed using similar mathematical calculations. The algorithm of the same is given below.

Algorithm:

*Input:*   Modified song signal with interchanged coefficients of specified harmonics.
*Output:* Modified song signal with embedded message.
*Method:* Extraction of coefficients and reallocating them in their original positions.

Step 1: Apply FFT over embedded signal (song) x(n,2) to get the magnitude values of frequencies and put into s(i) , 1<=i<=n. Find the coefficients(Ci) using equation  (3)  of some specified of harmonics. The selected harmonic number i is calculated by similar way as used in authentication algorithm.

Step 2:  Interchange the value of coefficients (each p and q) between two values of p, where i=1, 2, 3, … and i<=n as done in authentication algorithm.

Step 3: Apply IFFT over the output values s(i) , 1<=i<=n of step 2  to get the  sampled values  of original song signal with embedding secret message.

## 2.5. Emb-Extraction

The separation of embedding message is performed using simple way. The techniques are described below.

*Input*:   Modified song signal with embedded message.
*Output*:  Original song with separated embedded message.
*Method*:  Exacting sampled values and authenticating codes of original song.

Step 1:    Find FFT of output of extracted sample values and search the specified positions where we have inserted the message.

Step 2:    Collect all non-zero magnitudes values form above position.

Step 3:    Represent the collected magnitude values into equivalent small bit sequences as value 0 by 00, 0.0001 by 01, 0.0010 by 10, and 0.0011 by 11 respectively. Therefore, the ASCII bit sequence of input message is the sequence of all small bit sequences putting side by side. After getting the ASCII bit sequence, we can easily get the respective hidden encrypted message which can be verified with the encrypted string of original message by MD5 algorithm.

## 3. EXPERIMENTAL RESULTS

Encoding and decoding technique have been applied over 10 seconds recorded song, the song is represented by complete procedure along with results in each intermediate step has been outlined in subsections 3.1.1 to 3.1.4. The results are discussed in two sections out of which 3.1 deals with result associated with MDAC and that of 3.2 gives a comparison with existing techniques.

## 3.1. Results

For experimental observation, strip of 10 seconds classical song ('100 Miles From Memphis', sang by Sheryl Crow) has been taken. The sampled value of the song is given in table 1 as a two dimensional matrix. Figure 2 shows amplitude-time graph of the original signal. MDAC is applied on this signal and as a first step of the procedure which is performed FFT over input song signal. The output generated in the process is shown in figure 3 (number of sampled values is 441000). Selected coefficient values are shown in figure 4. Figure 5 shows the difference of frequency ratio of original and interchanged coefficients of selected harmonics. From figure 5, it is seen that the deviation is very less and will not affect the quality of the song at all.

### 3.1.1. Original Recorded Song Signals (10 seconds)

The values for sampled data array x(n,2) from the original song is given in table 1. Whereas the graphical representation of the original song, considering all rows (441000) of x(n,2) is given in the figure 2.

Table 1. Sampled data array x(n,2).

| Sl no | x(k,1) | x(k,2) |
|---|---|---|
| … | … | … |
|  | 0 | 0.0001 |
|  | 0.0000 | 0.0000 |
|  | -0.0009 | -0.0009 |
|  | -0.0006 | -0.0007 |
|  | -0.0012 | -0.0012 |
|  | -0.0014 | -0.0014 |
|  | -0.0016 | -0.0017 |
|  | -0.0023 | -0.0022 |
|  | -0.0027 | -0.0027 |
|  | -0.0022 | -0.0021 |
| … | … | … |

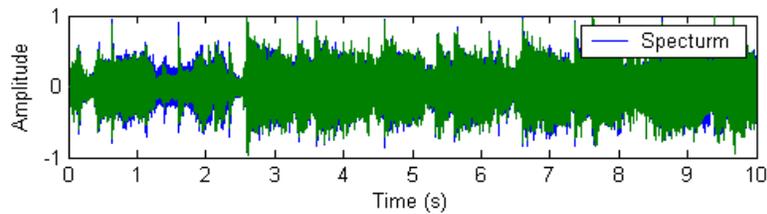

Figure 2. Original song ('100 Miles from Memphis', sang by Sheryl Crow)

### 3.1.2. Modified Song after Alternating Coefficient Values and Embedding Secret Key (10 seconds)

The graphical representation of the modified song signal is shown in the figure 3.

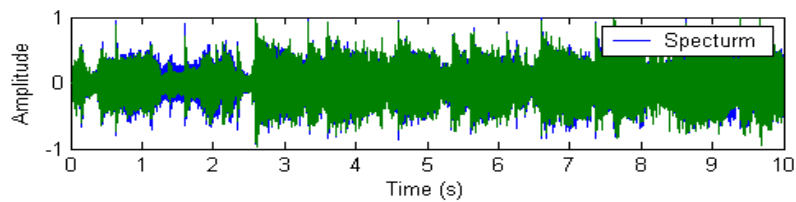

Figure 3. Modified song after alternating coefficient values and embedding secret key

### 3.1.3. Coefficient Values of Selected Harmonics

The graphical representation of the selected harmonics is shown in the figure 2.

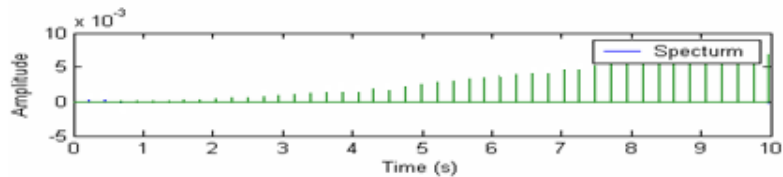

Figure 4. Selected coefficient values

### 3.1.4. The Difference of Magnitude Values between Original Signal and Modified signal

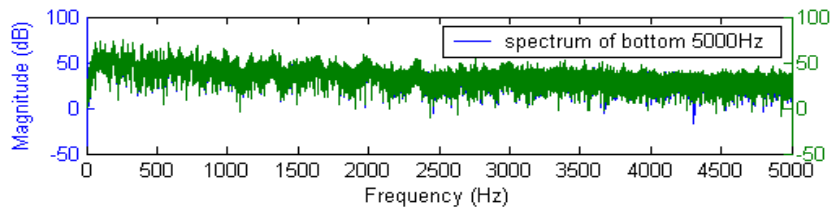

Figure 5. The magnitude values difference between signal figure 2 and figure 3.

## 3.2. Comparison with Existing Systems

Various algorithms [5] are available for embedding information with audio signals. They usually do not care about the quality of audio but we are enforcing our authentication technique without changing the quality of song. A comparison study of properties of our proposed method with Data hiding via phase manipulation of audio signals (DHPMA) [3] and A Fourier Transform Based Authentication of Audio Signals through Alternation of Coefficients of Harmonics (FTAT) [7] before and after embedding secret message/modifying parts of signal (16-bit stereo audio signals sampled at 44.1 kHz.) are given in table 2, table3 and table4. In FTAT, coefficient alternation has been taken place in one dimensional way, only between closely harmonics. Therefore, a linear alternation of coefficients of harmonics was used as a security measure. In the present paper (MDAC), alternation of coefficient are taken place between selected harmonics by a specific number which is provided by a sine wave as describe above [step 2 of authentication algorithm]. Therefore, it provides double security measure with alternation of coefficients and this technique is being applied in higher frequency region (above 20,000 Hz) which is beyond the audible range, i.e., it has less affect over audible quality of song signal over FTAT technique. On the other hand, the message hiding is performed in the form of generated string of message digest algorithm (AD5) which accepts variable size of input data and generates unique string , i.e., changing one part/bit may change the total message. Different statistical parameters like AD, NAD, MSE, NMSE, SNR and PSNR may use to measure the quality of audio signal. Average absolute difference (AD) is used as the dissimilarity measurement between original song and modified song to justify the modified song. Whereas a lower value of AD signifies lesser error in the modified song. Normalized average absolute difference (NAD) is quantization error is to measure normalized distance to a range between 0 and 1. Mean square error (MSE) is the cumulative squared error between the embedded song and the original song. A lower value of MSE signifies lesser error in the embedded song. The SNR is used to measure how much a signal has been tainted by noise. It represents embedding errors between original song and modified song and calculated as the ratio of signal power (original song) to the noise power corrupting the signal. A ratio higher than 1:1 indicates more signal than noise. The PSNR is often used to assess the quality measurement between the original and a modified song. The higher the PSNR represents the better the quality of the modified song. Thus from our experimental results of benchmarking parameters (NAD, MSE, NMSE, SNR and PSNR) in proposed method obtain better performances without affecting the audio quality of song.

Table 3 gives the experimental results in terms of SNR (Signal to Noise Ratio) and PSNR (Peak signal to Noise Ratio). Table 4 represents comparative values of Normalized Cross-Correlation (NC) and Correlation Quality (QC) of proposed algorithm with DHPMA. The Table 5 shows PSNR, SNR, BER (Bit Error Rate) and MOS (Mean opinion score) values for the proposed algorithm. Here all the BER values are 0. The figure 6 summarizes the results of this experimental test. It shows this algorithm's performance is stable for different types of audio signals.

Table 2. Metric for different distortions

| Sl No | Statistical parameters for Differential distortion | Value using MDAC | Value using FTAT | Value using DHPMA |
|---|---|---|---|---|
| 1 | MD | 0.0031 | 0.4456 | 3.6621e-004 |
| 2 | AD | 0.0015 | 2.5590e-005 | 2.0886e-005 |
| 3 | NAD | 0.0106 | 1.7576e-004 | 0.0063 |
| 4 | MSE | 3.0486e-006 | 5.8431e-006 | 1.4671e-009 |
| 5 | NMSE | 1.1834e+004 | 6.1743e+003 | 8.4137e-005 |

Table 3. SNR and PSNR

| Sl No | Statistical parameters for Differential distortion | Value using MDAC | Value using FTAT | Value using DHPMA |
|---|---|---|---|---|
| 1 | Signal to Noise Ratio (SNR) | 36.1023 | 37.9059 | 40.7501 |
| 2 | Peak Signal to Noise Ratio (PSNR) | 54.1203 | 52.2330 | 45.4226 |

Table 4. Representing NC and QC

| Sl No | Statistical parameters for Correlation distortion | Value using MDAC | Value using FTAT | Value using DHPMA |
|---|---|---|---|---|
| 1 | Normalised Cross-Correlation(NC) | 1 | 1 | 1 |
| 2 | Correlation Quality (QC) | -.0780 | -0.0803 | -0.5038 |

Table 5. Showing SNR, PSNR BER, MOS

| Audio (1s) | SNR | PSNR | BER | MOS |
|---|---|---|---|---|
| Song1 | 39.6410 | 54.3922 | 0 | 5 |
| Song2 | 36.3916 | 47.8229 | 0 | 5 |
| Song3 | 37.7033 | 49.6415 | 0 | 5 |
| Song4 | 38.9049 | 53.2230 | 0 | 5 |
| Song5 | 36.7891 | 50.1851 | 0 | 5 |

This quality rating (Mean opinion score) is computed by using equation (4).

$$Quality = \frac{5}{1 + N * SNR} \quad (4)$$

where N is a normalization constant and SNR is the measured signal to noise ratio. The ITU-R Rec. 500 quality rating is perfectly suited for this task, as it gives a quality rating on a scale of 1 to 5 [6]. Table 6 shows the rating scale, along with the quality level being represented.

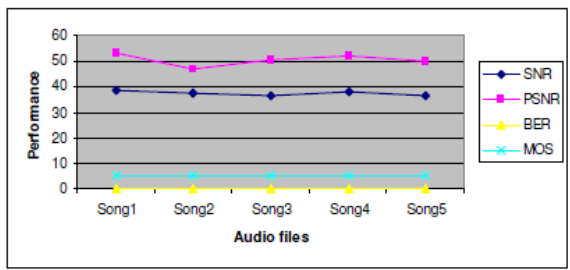

Figure 6. Performance for different audio signals.

Table 6. SNR and PSNR

| Rating | Impairment | Quality |
|---|---|---|
| 5 | Imperceptible | Excellent |
| 4 | Perceptible, not annoying | Good |
| 3 | Slightly annoying | Fair |
| 2 | Annoying | Poor |
| 1 | Very annoying | Bad |

## 4. CONCLUSION AND FUTURE WORK

In this paper, an algorithm for alternating closely coefficients of harmonics in higher frequency region with embedding secret key with the help of message digest method over song signal has been proposed which will not affect the audible quality of song signal but it will ensure the detection the distortion of song signal characteristics. The present algorithm is also very easy to implement.

This technique is developed based on the observation of characteristics of different songs but the mathematical model for representing the variation of those characteristics after modification may be formulated in future. It also can be extended to embed an image/audio into an audio signal instead of text message. The perfect estimation of percentage of threshold numbers of sample data of song that can be allow to change for a normal conditions will be done in future with all possibilities of errors.

## Authors

Uttam Kr. Mondal, has received his Bachelor of Engineering (B.E) degree in Information Technology in 2004 and Master of Technology (M.Tech) in Information Technology in 2006 from University of Calcutta, India. He has now working as an Asst. Professor in department of Computer Science & Engineering and Information Technology in College of Engg. & Management, Kalaghat, West Bengal, India. His research areas include cryptography & Network Security, Audio signal authentication. He has 15 publications in National and International conference proceedings and journal.

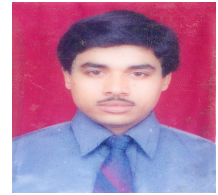

Jyotsna Kumar Mandal, M. Tech.(Computer Science, University of Calcutta),Ph.D.(Engg., Jadavpur University) in the field of Data Com pression and Error Correction Techniques, Professor in Computer Science and Engineering, University of Kalyani, India. Life Member of Computer Society of India since 1992 and life member of cryptology Research Society of India. Dean Faculty of Engineering, Technology & Management, working in the field of Network Security, Steganography, Remote Sensing  & GIS Application, Image Processing. 25 years of teaching and research experiences. Eight Scholars awarded Ph.D. and 8 are pursuing.  Total number of publications 189.

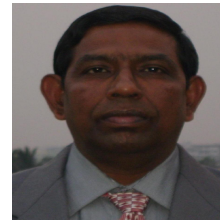